\documentclass{emulateapj}

\begin{document}

\shortauthors{Luhman et al.}
\shorttitle{New Stars and Brown Dwarfs in Upper Scorpius}

\title{New Young Stars and Brown Dwarfs in the Upper Scorpius Association\altaffilmark{1}}

\author{
K. L. Luhman\altaffilmark{2,3},
K. A. Herrmann\altaffilmark{4},
E. E. Mamajek\altaffilmark{5,6},
T. L. Esplin\altaffilmark{7},
and M. J. Pecaut\altaffilmark{8}
}

\altaffiltext{1}
{Based on observations made with the Two Micron All Sky Survey,
the Deep Near-Infrared Survey of the Southern Sky,
the {\it Wide-field Infrared Survey Explorer},
the United Kingdom Infrared Telescope Infrared Deep Sky Survey, 
the Visible and Infrared Survey Telescope for Astronomy Hemisphere Survey,
Pan-STARRS1, the {\it Gaia} mission, the NASA Infrared Telescope Facility,
Cerro Tololo Inter-American Observatory, and Gemini Observatory.}

\altaffiltext{2}{Department of Astronomy and Astrophysics,
The Pennsylvania State University, University Park, PA 16802, USA;
kluhman@astro.psu.edu}
\altaffiltext{3}{Center for Exoplanets and Habitable Worlds, The
Pennsylvania State University, University Park, PA 16802, USA}
\altaffiltext{4}{The Pennsylvania State University,
1 Campus Drive, Mont Alto, PA 17237, USA}
\altaffiltext{5}{Jet Propulsion Laboratory, California Institute of Technology,
4800 Oak Grove Dr., Pasadena, CA 91109, USA}
\altaffiltext{6}{Department of Physics \& Astronomy, University of Rochester, 
500 Wilson Blvd., Rochester, NY 14627, USA}
\altaffiltext{7}{Steward Observatory, University of Arizona, 933 North Cherry 
Avenue, Tucson, AZ 85721, USA}
\altaffiltext{8}{Department of Physics, Rockhurst University, 1100 Rockhurst 
Rd, Kansas City, MO 64110-2508, USA}

\begin{abstract}

To improve the census of the Upper Sco association ($\sim$11~Myr, $\sim$145~pc),
we have identified candidate members using parallaxes, proper motions,
and color-magnitude diagrams from several wide-field imaging surveys
and have obtained optical and infrared spectra of several hundred candidates
to measure their spectral types and assess their membership. We also
have performed spectroscopy on a smaller sample of previously known or
suspected members to refine their spectral types and evidence
of membership. We have classified 530 targets as members of Upper Sco,
377 of which lack previous spectroscopy. Our new compilation of all known
members of the association contains 1631 objects.
Although the census of Upper Sco has expanded significantly over the last
decade, there remain hundreds of candidates that lack spectroscopy.
The precise parallaxes and proper motions from the second data release
of {\it Gaia} should extend down to substellar masses in Upper Sco, which will
greatly facilitate the identification of the undiscovered members.

\end{abstract}


\section{Introduction}
\label{sec:intro}

The Upper Scorpius subgroup in the Scorpius-Centaurus OB association is located 
at a distance of $\sim145$~pc, has an age of $\sim$11~Myr, and contains
$>$1000 members \citep{pm08,pec12}, making it the largest stellar population at
$\lesssim10$~Myr within 300~pc.
As such, Upper Sco offers some of the best available
constraints on statistical properties of young stellar populations, such as
their initial mass function (IMF), stellar multiplicity, planet occurrence
rate, and disk fraction.
The association is old enough that its natal cloud has dispersed,
so its members have only modest extinction ($A_V\lesssim3$) and can be
observed at optical wavelengths.
However, since the members are distributed across a large area of sky
($\sim100$~deg$^2$), it has been challenging to obtain a complete
census of the association.
Surveys for members of Upper Sco have utilized wide-field imaging at X-ray, 
ultraviolet, optical, and infrared (IR) wavelengths
\citep{wal94,mar98b,mar98a,mar04,mar10,pre98,pre01,pre02,dez99,ard00,hoo00,sle06,sle08,tor06,lod06,lod07,lod08,lod11a,lod11b,lod13d,lod13b,lod18,lod13c,fin10,daw11,daw14,riz11,riz15,pec16,pen16,coo17,wil18}.
Members uncovered through that work have served as targets 
in surveys to find substellar companions through direct imaging
\citep{kra05,kra08,kra07,kra09,bej08,bil11,ire11,all13b,laf10,laf11,laf14,hin15,uya17},
detect and characterize circumstellar disks via IR and millimeter observations
\citep{rie05,chen05,chen11,car06,car08,car09,car14,pad06,sch07,dah09,ria09,ria12,luh12usco,mat12,mat13,riz12,riz15,daw13,jan15,bar16,bar17},
and measure photometric variability from a variety of phenomena
(e.g., eclipsing binaries, transiting planets) with {\it Kepler}'s K2 mission
\citep{kra15,alo15,lod15,rip15,sch15,ans16,dav16a,dav16b,dav17,man16,sca16,cod17,riz17,sta17,sta18,cod18,hed18,reb18}
as well as the Super Wide Angle Search for Planets \citep{mel17}
and the Kilodegree Extremely Little Telescope \citep{ans18}.

Although extensive work has been done in identifying the members of Upper Sco,
the current census likely has substantial incompleteness, particularly among
low-mass stars and brown dwarfs.
To improve the completeness of the census in Upper Sco, we are performing
a search for new members using photometry and astrometry from optical
and IR wide-field surveys, which include
the Two Micron All Sky Survey \citep[2MASS,][]{skr06},
the Deep Near-Infrared Survey of the Southern Sky \citep[DENIS,][]{epc99},
the United Kingdom Infrared Telescope Infrared Deep Sky Survey
\citep[UKIDSS,][]{law07}, the {\it Wide-field Infrared Survey Explorer}
\citep[{\it WISE},][]{wri10}, Pan-STARRS1 \citep[PS1,][]{kai02,kai10},
the {\it Gaia} mission \citep{per01,deb12}, and the Visible and Infrared Survey 
Telescope for Astronomy (VISTA) Hemisphere Survey \citep[VHS,][]{mcm13}.
In this paper, we present the new members found to date through our survey and
a compilation of all known members of Upper Sco.

\section{Known Members of Upper Sco}

During a study of circumstellar disks in Upper Sco,
\citet{luh12usco} compiled a list of known members of the association.
That list contained 863 sources, 181 of which comprised an initial 
sample of new members found through the survey that we are presenting
in this work. Many additional members have been identified by subsequent
surveys of Upper Sco. We have revised and expanded the compilation of
members from \citet{luh12usco} to account for those surveys and new
membership constraints, as summarized in this section.

To compile known members of Upper Sco, we began by considering the 
120 stars (mostly B through F types) that were identified as probable members
by \citet{dez99} based on proper motions and parallaxes from the 
{\it Hipparcos} mission.
We have excluded eight candidates that have been spectroscopically classified
as field stars \citep{hou82,hou88,chen11,pec12}, which consist of HIP~75916,
HIP~78271, HIP~79247, HIP~79258, HIP~79596, HIP~82140, HIP~83456, and HIP~83542.
The first data release of the {\it Gaia} mission \citep{gaia16a,gaia16b}
has provided new parallax measurements for 76 of the 112 remaining candidates.
Several stars exhibit discrepant {\it Gaia} parallaxes relative to the bulk
of the sample.
Among those stars, we assigned membership to those that 
are known or suspected to be binaries (HIP~78233, HIP~78581, HIP~80896) since
the presence of a close companion can lead to erroneous astrometry.
Another candidate with a discrepant parallax, HIP~82218, is treated as a
member because of the abundance of evidence of its membership
\citep{chen11}. We reject HIP~78963, HIP~79860, HIP~79987, and HIP~80586
based on their parallaxes \citep[and radial velocity for HIP~80586,][]{nor04}.

\begin{figure}
\epsscale{1}
\plotone{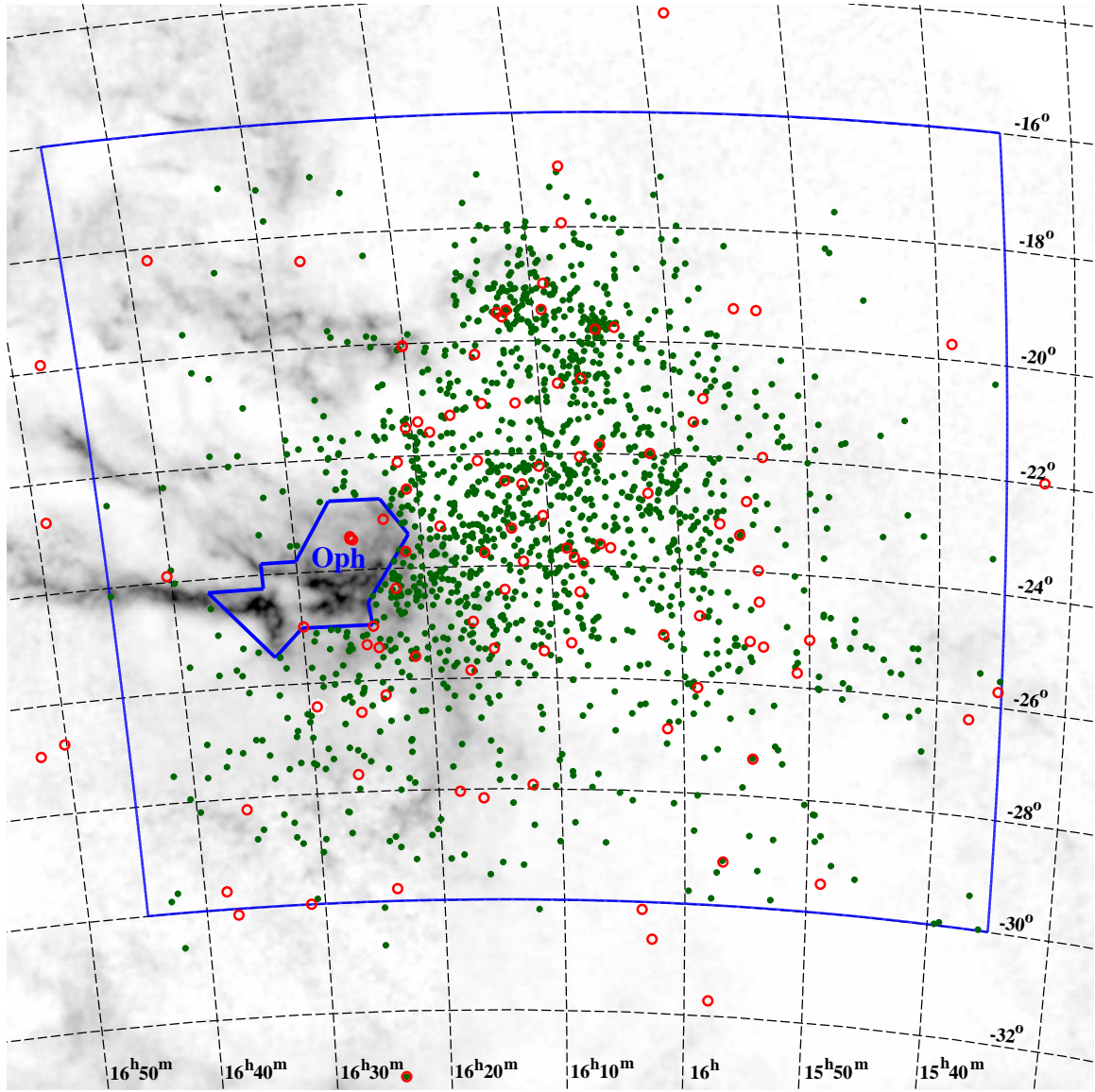}
\caption{
Spatial distribution of stars identified as probable members of Upper Sco
by \citet{dez99} using {\it Hipparcos} astrometry (open circles),
excluding 12 stars that are nonmembers based on spectral classifications
or {\it Gaia} parallaxes. We also include all other stars that we have adopted
as members of the association (filled circles, Table~\ref{tab:mem}).
In our survey for new members of Upper Sco, we have considered the area from
$\alpha=15^{\rm h}35^{\rm m}$ to $16^{\rm h}45^{\rm m}$ and 
$\delta=-30$ to $-16\arcdeg$ (rectangle) and outside of the Ophiuchus
region (small polygon).
The Ophiuchus dark clouds and the diffuse clouds across portions of Upper Sco
are displayed with a map of extinction \citep[gray scale,][]{dob05}.
}
\label{fig:map1}
\end{figure}

The positions of the 108 unrejected stars from \citet{dez99} are
plotted on the map in Figure~\ref{fig:map1}. Based on the distribution
of those stars, we selected $15^{\rm h}35^{\rm m}$ and $16^{\rm h}45^{\rm m}$
in right ascension ($\alpha$) and $-30\arcdeg$ to $-16\arcdeg$ in
declination ($\delta$) as the boundaries for our survey, which encompass
97 of the 108 probable members. In Figure~\ref{fig:map1}, we have
marked our adopted boundary between Upper Sco and the stellar population
associated with the Ophiuchus dark clouds, which was defined by \citet{esp18}.
Five of the stars from \citet{dez99} appear within that boundary for
Ophiuchus, and hence are excluded from our membership list for Upper Sco.
The 103 remaining stars outside of Ophiuchus are included in the list.

We have compiled additional members of Upper Sco from the previous surveys
cited in Section~\ref{sec:intro}, a survey for disk-bearing members by
\citet{esp18}, a proper motion survey of Ophiuchus that extends into Upper
Sco (T. Esplin, in preparation), and the survey that we are presenting.
We have also considered young stars that have been found in the vicinity
of Upper Sco through studies that were not explicitly searching for members
of the association \citep[e.g.,][]{her72,giz02,all06,shk09,kir10,rom12,bes17}.
We have adopted stars as members if they are located between the boundary
of Ophiuchus and the outer boundary of our survey
field (Figure~\ref{fig:map1}), have measured spectral types, exhibit evidence 
of youth, and have parallaxes or proper motions that are similar to those of
the adopted members from \citet{dez99} when available.
We have included a few additional stars outside of our survey field that have
been adopted as members in multiple previous studies (see 
Figure~\ref{fig:map1}).
Diagnostics of youth consist of Li~I absorption, gravity-sensitive
absorption features (e.g., Na~I, FeH, H$_2$O), and mid-IR excess emission
that is indicative of a circumstellar disk. The first two diagnostics are
not applicable to B and A stars, and many members of young populations lack
disks, so we do not require evidence of youth for members at those types.
The membership classification of a star can change as new data become
available \citep[e.g.,][]{muz03,bou09}, so some stars previously identified
as members do not appear in our compilation and some stars previously
classified as field stars are now adopted as members. We have attempted
to consider all available membership constraints for each candidate that
we have examined.

In Table~\ref{tab:mem}, we list the stars that we have adopted as known members
of Upper Sco. We have included the available measurements of spectral types,
the types that we adopt, and our estimates of extinction 
(Section~\ref{sec:class}). This compilation contains 1631 objects.
Spectra have been obtained for 185, 31, and 530 members by
\citet{esp18}, T. Esplin (in preparation), and this work, respectively,
166, 31, and 377 of which have not been observed with spectroscopy previously.

\section{Identification of Candidate Members}

\subsection{Sources of Data}
\label{sec:sources}

Since the members of Upper Sco are distributed across a large area of sky,
we have used wide-field imaging surveys to search for members of the
association. The surveys have provided photometry in several optical and IR
bands: $JHK_s$ from the 2MASS Point Source Catalog,
$i$ from the third data release of DENIS,
$ZYJHK$ from the science verification release and data release 10 of UKIDSS,
$W1$ and $W2$\footnote{{\it WISE} obtained images in bands at
3.5, 4.5, 12, and 22~$\mu$m, which are denoted as $W1$, $W2$, $W3$, and $W4$,
respectively.} from the AllWISE Source Catalog,
$YJHK$ from the fifth data release of VISTA VHS,
$G$ from the first data release of {\it Gaia}, and
$rizy_{P1}$ \citep{ton12} from the first data release of the 3$\pi$ survey
from PS1 \citep{cha16,fle16}.
Additional bands are available from some of these catalogs, but they
are redundant with data adopted from other surveys. Some of the catalogs
provide multiple methods of measuring photometry for a given band.
We adopted the $1\arcsec$ radius aperture magnitudes from UKIDSS and VHS and
the point-spread-function magnitudes from the stacked images in PS1.
To avoid saturated measurements, we excluded PS1 data brighter than
15~mag, UKIDSS data at $Z<11.4$, $Y<11.3$, $J<10.5$, $H<10.2$, and $K<9.7$,
and VHS data at 2MASS magnitudes of $J<13$, $H<13$, and $K_s<12.5$.
We also omitted $Y$ data from VHS for $J<13$.
When photometry was available at two epochs in $K$ from UKIDSS,
we adopted the average of those measurements.
For some of the strips in the DENIS catalog, all of the measurements exhibited 
unusually large photometric errors. We did not use data from those strips.
In addition to the above photometry, we have utilized proper motions and
parallaxes from {\it Gaia} DR1 and proper motions from UKIDSS, the fifth
release of the U.S. Naval Observatory CCD Astrograph Catalog 
\citep[UCAC5,][]{zac17}, and the {\it Gaia}-PS1-Sloan Digital Sky Survey (GPS1) 
Proper Motion Catalog \citep{tia17}\footnote{In addition to the databases
listed in its name, GPS1 also employed astrometry from 2MASS.}.

We combined the above catalogs by first identifying all matching sources.
To merge the photometry from different surveys, we treated the following
filters as the same: $JHK_s$(2MASS)/$JHK$(UKIDSS)/$JHK$(VHS) and
$Y$(UKIDSS)/$Y$(VHS).
For $JHK_s$, we adopted 2MASS measurements if the errors were $\leq$0.06.
When the 2MASS errors were larger than that threshold, we adopted the
measurements with the smallest errors from among 2MASS, UKIDSS, and VHS.
For $Y$, we used the measurements with the smallest errors from UKIDSS and VHS.
The bands $Z$(UKIDSS)/$z_{P1}$ and $Y$(UKIDSS)/$y_{P1}$ are sufficiently
different that they are used separately in our analysis.

\begin{figure}
\epsscale{1}
\plotone{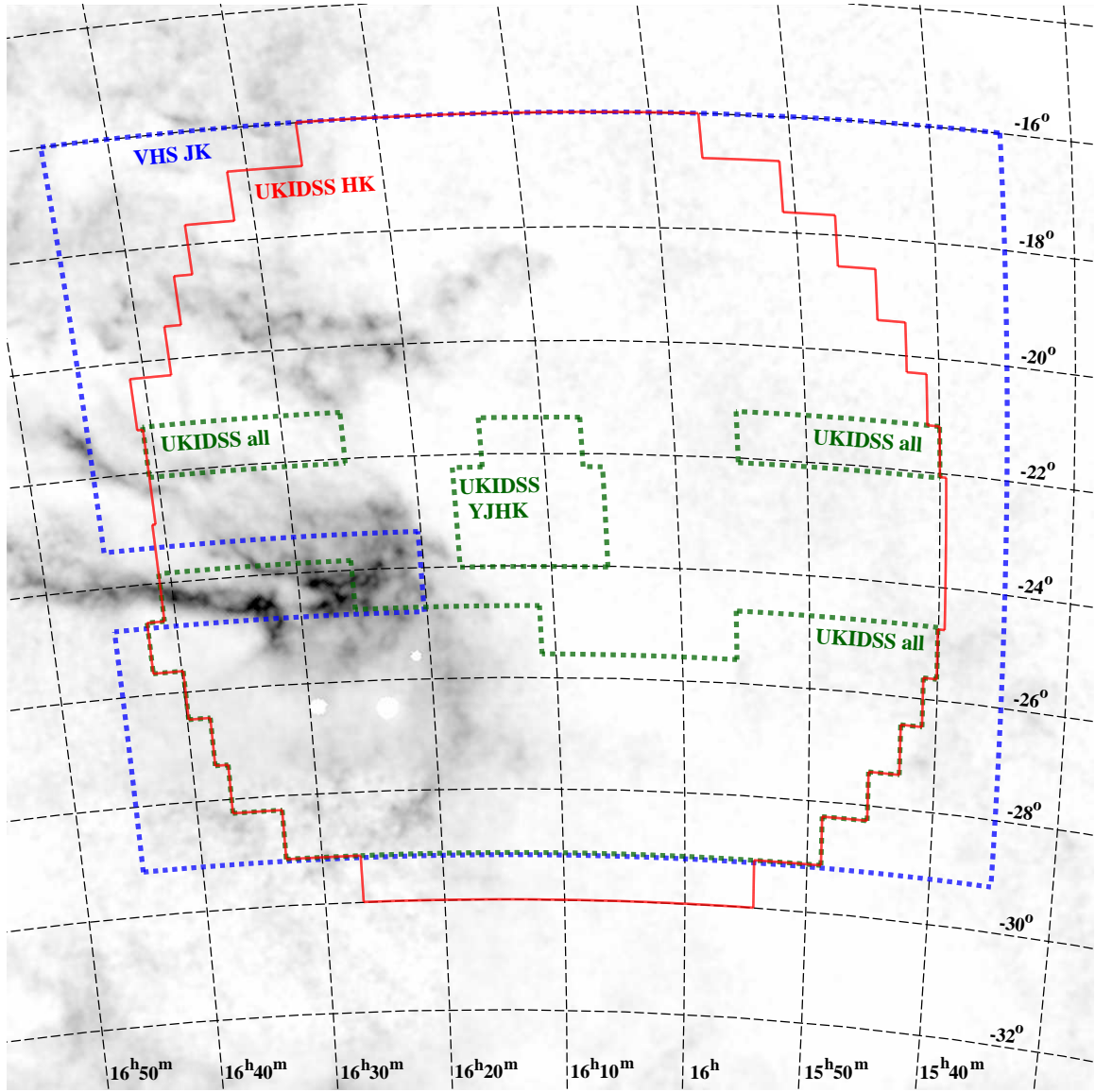}
\caption{
Areas in our survey field of Upper Sco that have been imaged by
UKIDSS and VISTA VHS. The clouds in Ophiuchus and Upper Sco 
are displayed with a map of extinction \citep[gray scale,][]{dob05}.
}
\label{fig:map2}
\end{figure}

The images from 2MASS, {\it WISE}, {\it Gaia}, and PS1 cover all or nearly
all of our survey field in Upper Sco. After excluding the strips with
large photometric errors, the remaining DENIS data covered $\sim85$\% of
the survey field. The boundaries of the areas imaged by UKIDSS and VHS
are indicated on the map in Figure~\ref{fig:map2}. Those areas were not
fully imaged by those surveys; small patches within them lack data.
The coverage of UKIDSS in Upper Sco was also presented by \citet{lod13c}.
Most of the available VHS data were obtained in $J$ and $K$.
Only a very small portion of Upper Sco has been imaged by VHS in $Z$
and $H$, which is not shown in Figure~\ref{fig:map2}.
Imaging has been performed in small areas of Upper Sco that is deeper than
the wide-field surveys cited above \citep{lod13b,lod18,pen16}.
Those data are not utilized in our survey since only the raw images are
publicly available.

\subsection{Proper Motions and Parallaxes}
\label{sec:pm}

We have used the surveys that we have compiled to identify
candidate members of Upper Sco via parallaxes, proper motions, and
color-magnitude diagrams (CMDs). We describe our selection criteria
for the parallaxes and proper motions in this section. Our analysis
of the CMDs is discussed in the next section.

Measurements of parallaxes are available from {\it Gaia} DR1 for 169
of the brightest stars in our compilation of known members of Upper Sco
($G\sim5$--11.5). The median value of these parallax data corresponds to 146~pc,
which agrees with previous distance estimates for the association \citep{pm08}.
Most of the parallaxes have 1~$\sigma$ errors that overlap with a distance
range of 125--165~pc, so we have adopted this criterion when identifying
candidate members among other stars that have parallaxes from {\it Gaia} DR1.

To select candidates based on proper motions, we have utilized the motions
measured by UKIDSS, GPS1, and UCAC5. The GPS1 catalog contains multiple options
for proper motion that differ in their fits or combinations of data.
Two of those options, the robust fit and cross-validated fit, use all
available data from {\it Gaia} DR1, PS1, 2MASS, and Sloan (the latter
does not cover Upper Sco). We compared the measurements from those two fits
for the known members of Upper Sco. The formal errors from the robust fit are
smaller than those from the cross-validated fit, but the two sets of data
exhibit similar dispersions, which would suggest that they have similar errors.
We have adopted the motions from the cross-validated fit since they contain
fewer outliers relative to the average motion of the members.
We also have computed one set of proper motions from a combination of
2MASS and {\it Gaia} positions and a second set of motions from
a combination of 2MASS and PS1 positions. The PS1 data were based on
coadditions of images from multiple epochs.
In our analysis, we have excluded proper motion measurements
with errors $>$10~mas~yr$^{-1}$ in $\mu_\alpha$ or $\mu_\delta$.
The numbers of known members with proper motion measurements are
169 for {\it Gaia} ($G\sim5$--11.5),
929 for GPS1 ($G\sim12.5$--18.5),
823 for UCAC5 ($G\sim4.5$--15.5),
1297 for 2MASS/{\it Gaia} ($G\sim5$--21),
895 for 2MASS/PS1 ($G\gtrsim13$), and
382 for UKIDSS ($G\gtrsim16$).
2MASS/PS1 and UKIDSS provide motions for a few hundred members that are
below the detection limit of {\it Gaia} ($G\gtrsim21$).

\begin{figure}
\epsscale{1.2}
\plotone{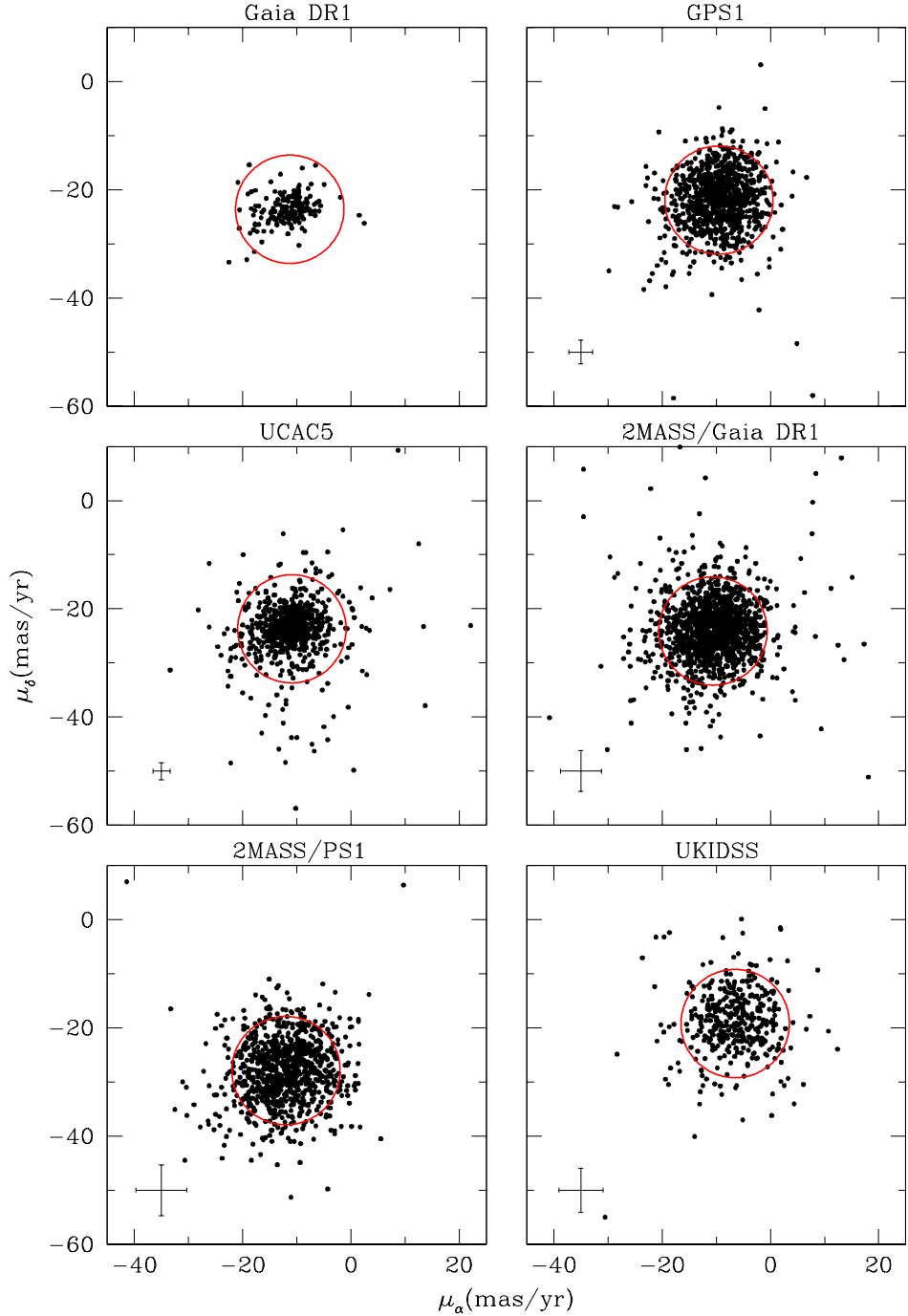}
\caption{
Proper motions for known members of Upper Sco from {\it Gaia} DR1,
GPS1, UCAC5, 2MASS/{\it Gaia} DR1, 2MASS/PS1, and UKIDSS.
Most of the outliers are known or suspected binaries, which can lead
to erroneous proper motion measurements, or have proper motions from other
sources that are consistent with membership.
For a given set of data, the large circle represents the threshold that
we have used for identifying new candidate members (see Section~\ref{sec:pm}).
The typical errors are indicated in the corner of each diagram except for
{\it Gaia} DR1, for which most errors are similar to the sizes of the points.
}
\label{fig:pm}
\end{figure}

As with the parallaxes, we have used known members of Upper Sco to
design our proper motion criteria for selecting candidate members.
We began by estimating the intrinsic spread in motions within
Upper Sco using the most accurate measurements that are available, which
are those from {\it Gaia}. Those data are plotted in a diagram
of $\mu_\delta$ versus $\mu_\alpha$ in Figure~\ref{fig:pm}.
They exhibit a dispersion ($\sigma\sim3$~mas~yr$^{-1}$) that is significantly
larger than that expected from measurement errors ($\sim0.3$~mas~yr$^{-1}$)
or projection effects that occur across the area covered by Upper Sco
(0.5--1~mas~yr$^{-1}$). Thus, it appears that the scatter is
dominated by the dispersion in kinematics and distances among the members
\citep[see also][]{wri18}, and thus reflects the intrinsic spread in
the proper motions of the association. Most of the known members with
{\it Gaia} motions are encompassed by a radius of 10~mas~yr$^{-1}$, so we
have adopted that threshold for our selection of candidates.
For each of the six sources of proper motions that we have utilized,
we identified stars that have 1~$\sigma$ errors that overlap
with a radius of 10~mas~yr$^{-1}$ from the median value of the motions
of the known members. We also identified stars that are beyond
the threshold by more than 2~$\sigma$. Measurements at 1--2~$\sigma$ are
ignored. The 10~mas~yr$^{-1}$ threshold is indicated in Figure~\ref{fig:pm}
for each source of proper motions. We classified each star as 
either a candidate member or a probable field star based on the
first data set in which it was flagged as within 1~$\sigma$ or beyond
2~$\sigma$ of Upper Sco from among {\it Gaia}, GPS1, UCAC5, 2MASS/{\it Gaia},
2MASS/PS1, and UKIDSS. Given this approach, a star with one proper motion
measurement that indicates nonmembership can be considered a candidate if
another measurement from a more preferred source supports membership.
A small number of stars in our compilation of known
members would be rejected by this criterion (i.e., they appear as outliers
in Figure~\ref{fig:pm}), but they have been retained as members
because they are known or suspected binaries, which can lead to erroneous
proper motion measurements, or they have proper motions from other
sources that are consistent with membership.

We note that \citet{coo17} measured a dispersion of $\sigma\sim8$~mas~yr$^{-1}$
in each component of proper motion for the known members using data from
GPS1, UCAC5, and \citet{alt17}. They concluded that their measurement
was dominated by the kinematic spread of the population.
However, as discussed earlier, the more accurate {\it Gaia} motions show
a significantly smaller dispersion ($\sigma\sim3$~mas~yr$^{-1}$),
indicating that the dispersion from \citet{coo17} primarily reflected
measurement errors.

\subsection{Color-Magnitude Diagrams}

We have constructed CMDs for our survey field in Upper Sco using
the photometry described in Section~\ref{sec:sources}.
For these diagrams, we have selected $K_s$ (or $K$) as the vertical
axis and colors that contain this band because it offers the greatest
sensitivity to low-mass members of Upper Sco among the available bands.
For colors, that band is paired with $G$, $i$ from DENIS, $Z$ and $Y$
from UKIDSS, $i_{P1}$, $z_{P1}$, $y_{P1}$, and $H$. We also have included
a diagram with $W1-W2$.
Although the natal molecular cloud of Upper Sco has dispersed, non-negligible
dust extinction ($A_V\sim1$--3) is present across portions of the 
association, particularly near the Ophiuchus clouds (Figure~\ref{fig:map1}),
which increases contamination of background stars in regions of CMDs that are
inhabited by association members.
To reduce this contamination, we have estimated extinctions for individual
stars in the manner described by \citet{esp17tau} \citep[see also][]{luh03}
and have dereddened their positions in most of the CMDs accordingly.
The one exception is the diagram of $K_s$ versus $H-K_s$, where we use
the observed colors.
In this analysis, we have adopted $A_H/A_K=1.55$ \citep{ind05} and the
reddening relations from \citet{sch16},
\citet[][references therein]{xue16}, and \citet{dan18}, which produce 
$A_G/A_K=6.8$,
$A_{i(P1)}/A_K=6.51$,
$A_{z(P1)}/A_K=5.12$,
$A_{y(P1)}/A_K=4.17$,
$A_J/A_K=2.62$,
$A_{W1}/A_K=0.6$, and
$A_{W2}/A_K=0.48$.
For $i$ from DENIS and $Z$ and $Y$ from UKIDSS, we have estimated relations
of $A_i/A_K=6.1$, $A_Z/A_K=5$, and $A_Y/A_K=3.8$ based on the values of
$E(i-K_s)/E(J-H)$, $E(Z-K_s)/E(J-H)$, and $E(Y-K_s)/E(J-H)$ for reddened
stars in our survey field.

\begin{figure}
\epsscale{1.2}
\plotone{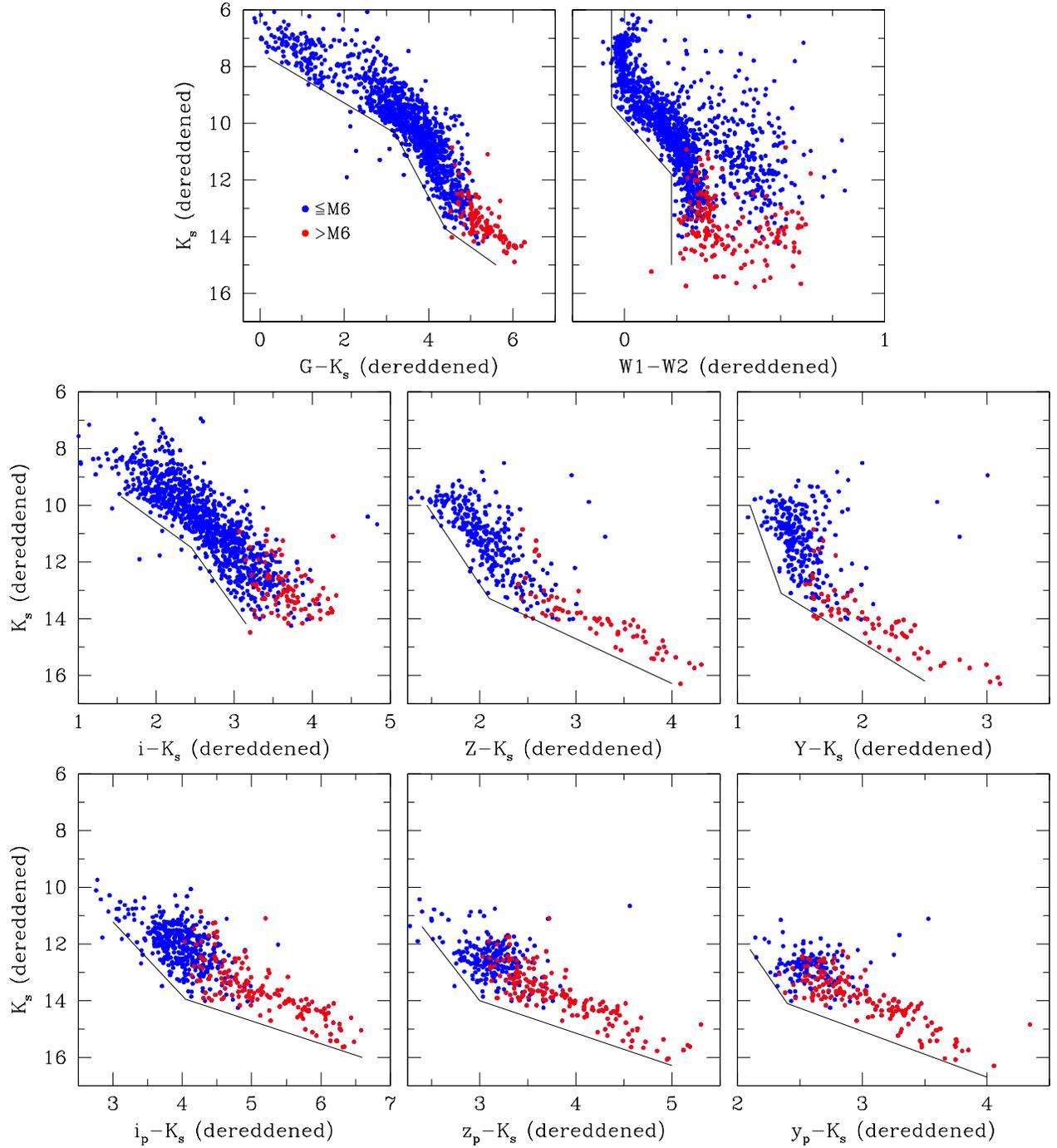}
\caption{
Extinction-corrected CMDs for the known members of
Upper Sco. These data are from 2MASS, DENIS, UKIDSS, PS1, VISTA VHS,
{\it WISE}, and {\it Gaia} (Section~\ref{sec:sources}). 
We have selected candidate members from the other stars detected in 
these surveys based on positions above the solid boundaries.
}
\label{fig:cmd}
\end{figure}

\begin{figure}
\epsscale{1.2}
\plotone{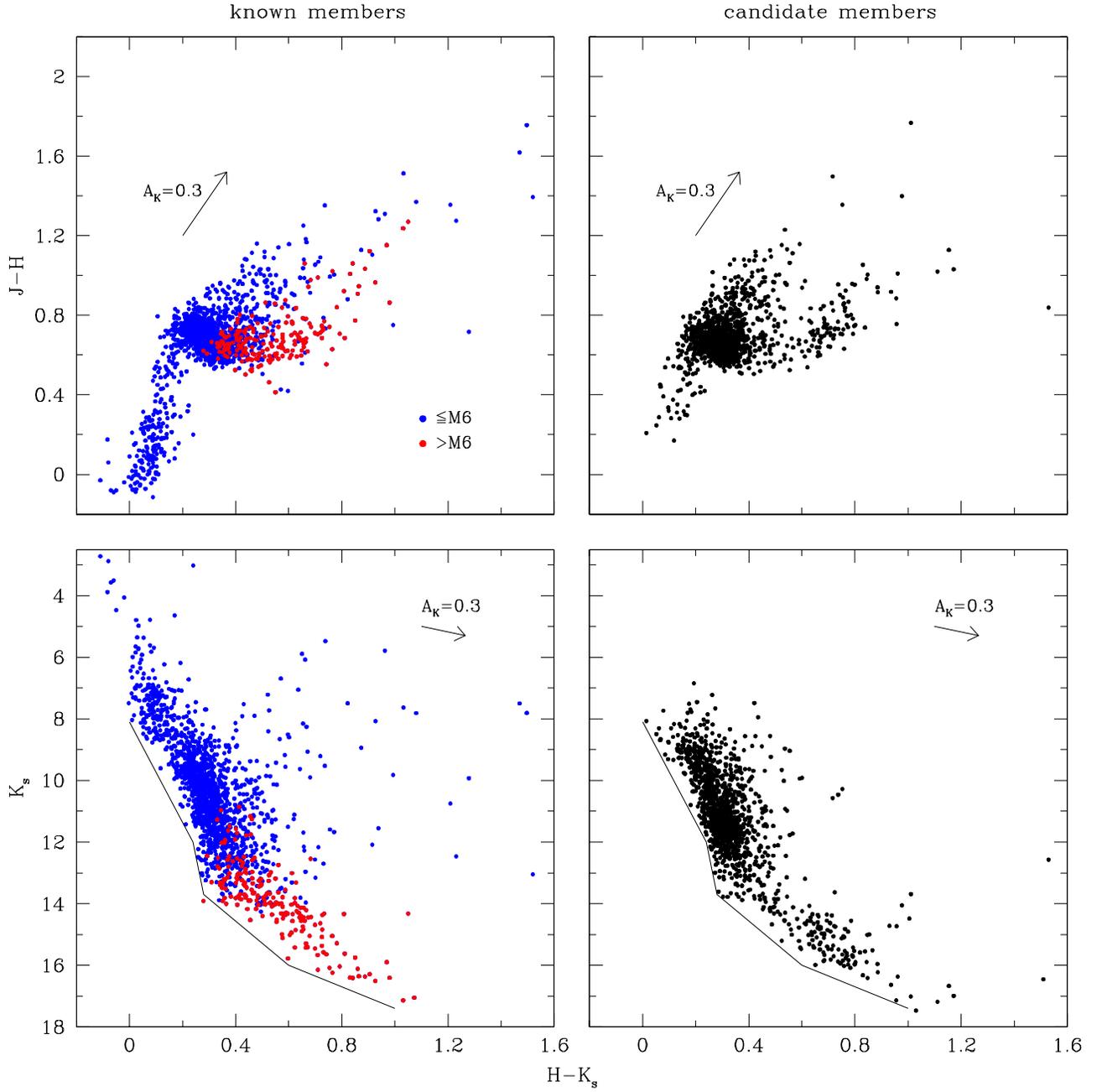}
\caption{
Near-IR color-color diagrams and CMDs for the known members of
Upper Sco (left, Table~\ref{tab:mem}) and candidate members identified with
proper motions and CMDs in Figures~\ref{fig:pm} and
\ref{fig:cmd} (right, Table~\ref{tab:cand}).
These data are from 2MASS, UKIDSS, and VISTA VHS.
We have rejected candidates from other diagrams that appear below the
solid boundary in $K_s$ versus $H-K_s$.}
\label{fig:cmd2}
\end{figure}

In Figure~\ref{fig:cmd}, we present the extinction-corrected CMDs for the
known members of Upper Sco, including those found in this work.
The CMD with the observed data in $H$ and $K_s$ is shown in Figure~\ref{fig:cmd2}.
In each diagram, we have defined a boundary that follows the lower edge
of the sequence of known members, which we have used for selecting candidate
members for spectroscopy. A few members appear below some of those boundaries;
most of them exhibit mid-IR excess emission that indicates the presence of
circumstellar material, so it is plausible that their low positions in the CMDs
are due to scattered light dominating their observed fluxes.

A source is considered a candidate member if it is selected with at least one
of the CMDs in Figure~\ref{fig:cmd}, is not rejected by any of those diagrams,
is a candidate in $K_s$ versus $H-K_s$ in Figure~\ref{fig:cmd2}, and is not a
probable field star based on the analysis of parallaxes and proper motions
in the previous section.
We inspected the resulting candidates in images from the employed surveys
and rejected those that appear to be galaxies.
In the next section, we describe spectroscopy of some of the candidates
that we have identified in this manner, which is used to measure their spectral
types and assess their membership.
Since we have added proper motions and CMDs to our selection process
as they have become available, some of the targets in our spectroscopic sample
that were observed early in our survey would now be rejected by our latest
criteria.  In Table~\ref{tab:cand}, we present the candidates that lack 
spectroscopy or that have insufficient information from the available spectra 
to determine membership. This sample contains 1196 objects.
We indicate in Table~\ref{tab:cand} the selection criteria that were satisfied
by each candidate. As noted in the previous section, it is possible for
a candidate to be rejected by a given proper motion measurement as long as
a more preferred source of proper motions favors membership.
To illustrate the distribution of magnitudes of these candidates, we have
plotted them in a diagram of $K_s$ versus $H-K_s$ in Figure~\ref{fig:cmd2}.
We have also included diagrams of $J-H$ versus $H-K_s$ for the known
members and the candidates. 
The candidates exhibit similar $J-H$ and $H-K_s$ colors as the known members.

\section{Spectroscopy of Candidates}

\subsection{Observations}

We have obtained optical and near-IR spectra of candidate members of Upper
Sco from the previous section to measure their spectral types and assess
their membership. We also have observed a sample of stars that were
previously known or suspected to be members to improve their spectral
classifications and evidence of membership. 
These data were taken between 2009 and 2017 at the
1.5~m SMARTS and 4~m Blanco telescopes at the Cerro Tololo Inter-American
Observatory (CTIO), the NASA Infrared Telescope Facility (IRTF),
the Southern Astrophysical Research Telescope (SOAR),
and the Gemini North and South telescopes.
We preferred optical spectroscopy for targets that were sufficiently bright
since the spectral types of young stars are defined at optical wavelengths
and Li absorption at 6707~\AA\ is a valuable diagnostic of youth.
For targets that were too faint for Li measurements or accurate optical
classifications, we employed near-IR spectroscopy instead.
The instrument configurations are summarized in Table~\ref{tab:log}. 
Spectroscopy was performed on 768 objects. The date and instrument for each
target are indicated in Table~\ref{tab:spec}.

During the several years of our spectroscopic survey, our selection criteria
were updated to incorporate new photometry and astrometry as they became
available. In addition, to utilize as many fibers as possible during the
multi-object observations with Hydra, we included targets that failed
some selection criteria. For these reasons, some of the spectroscopic
targets do not fully satisfy our latest set of criteria.

\begin{figure}
\epsscale{1}
\plotone{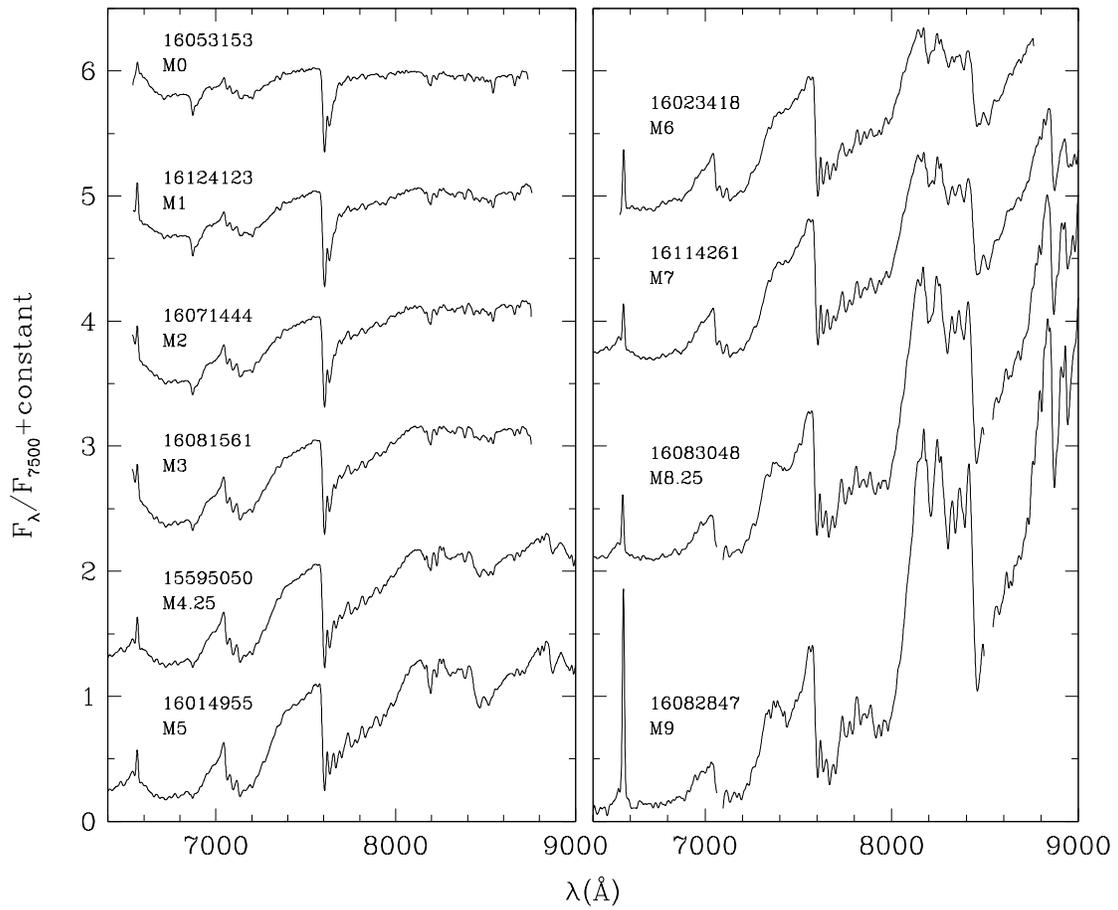}
\caption{
Examples of optical spectra of members of the Upper Sco association.
These data are displayed at a resolution of 13~\AA.
The data used to create this figure are available.
}
\label{fig:op}
\end{figure}

\begin{figure}
\epsscale{1.2}
\plotone{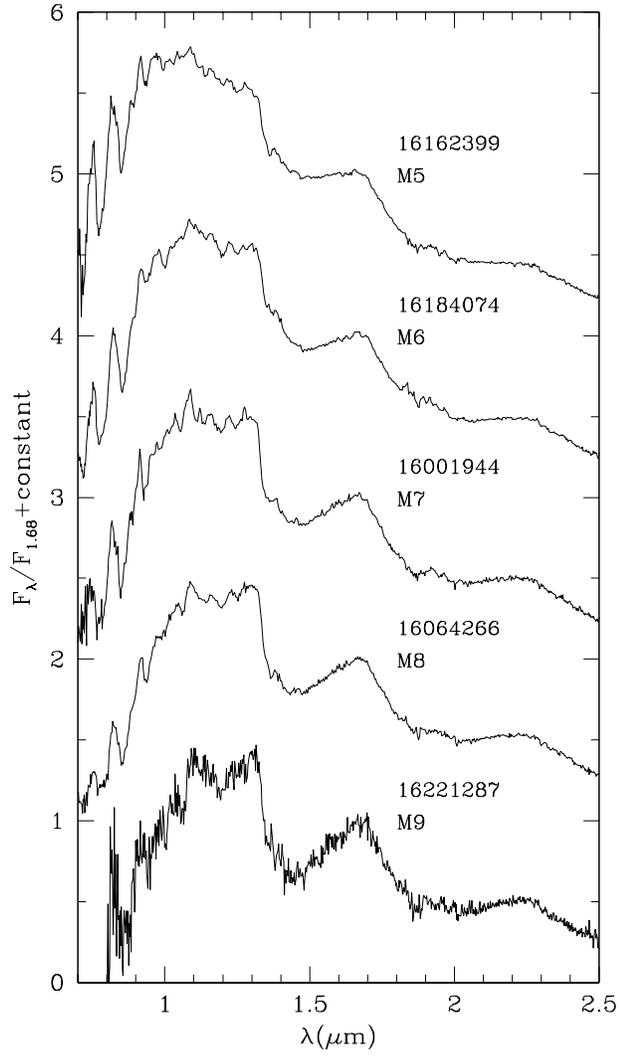}
\caption{
Examples of near-IR spectra of M-type members of the Upper Sco association.
They have been dereddened to match the slopes of the young standards from
\citet{luh17}. 
The data used to create this figure are available.
}
\label{fig:ir}
\end{figure}

The spectra from the RC spectrograph were reduced in the manner described
by \citet{pec16}. The SpeX data were reduced with the Spextool package 
\citep{cus04} and corrected for telluric absorption \citep{vac03}.
The spectra from the other instruments were reduced using routines within IRAF.
Examples of the reduced optical and IR spectra of young objects
are shown in Figures~\ref{fig:op} and \ref{fig:ir}, respectively.
All of the reduced spectra from our survey are provided in electronic
files that accompany those figures.
We have included the SpeX data for 2MASS J16183317-2517504~A and B
from \citet{luh05usco}. The slopes for the fiber spectra from Hydra are not 
reliable, particularly near the ends of the spectra. 
All of the slit observations were performed with the slit rotated to the
parallactic angle, so their spectral slopes should be accurate.
Many of the Goodman spectra contain significant fringing at redder wavelengths.
The spectra from the RC spectrograph were taken with two settings that
produced spectra that are separated by a gap in wavelength. The two spectra
for a given target are contained within a single file and have arbitrary
normalizations relative to each other.

\subsection{Classifications of New Spectra}
\label{sec:class}

To classify the optical and IR spectra that we have collected in Upper Sco,
we have applied the same methods that we have utilized in our previous surveys
of nearby star-forming regions.
We distinguished between young stars (which are likely to be members of the
association) and field dwarfs and giants using using Li absorption and
gravity-sensitive features (e.g, Na, H$_2$O), as available.
When present, IR excess emission and strong hydrogen emission indicate
the presence of a circumstellar accretion disk, providing additional
evidence of youth.
For targets that are classified as dwarfs or giants, we have measured
spectral types through comparison to the optical and IR spectra of standards
\citep{kee89,hen94,kir91,kir97,cus05,ray09}. We also classified the
optical and IR spectra of young stars at $<$M5 and $<$M0, respectively,
in the same way.
The optical spectra of young stars at M5--M9.5 were classified with
averages of the dwarf and giant standards \citep{luh97,luh98,luh99}.
We adopted the optical spectra of the youngest field L dwarfs from
\citet{cru07,cru09} and \citet{kir06,kir10} as standards for $>$M9.5
(spectral type suffixes of $\delta$ or $\gamma$, corresponding to
$\sim$10--100~Myr), although none of our optical spectra of young objects
exhibited types in that range. 
To classify the IR spectra of young stars at $\geq$M0, we applied the
standard spectra described by \citet{luh17}, which were measured for
optically-classified members of nearby star-forming
regions (1--10~Myr) at M0--M9.5, the youngest optically-classified dwarfs
in the field at L0, L2, and L4 \citep{kir06,kir10,cru07,cru09,cru18}, and
two members of the TW Hya association that have been previously assigned types
of L7 \citep{kel15,sch16b}.
By adopting IR standards of this kind for young objects, our classifications
should produce IR types that are on the same system as the optical types.
When comparing a target to a standard at a given spectral type, reddening was
artificially applied to the standard to match the spectral slope of the target.

\begin{figure}
\epsscale{1}
\plotone{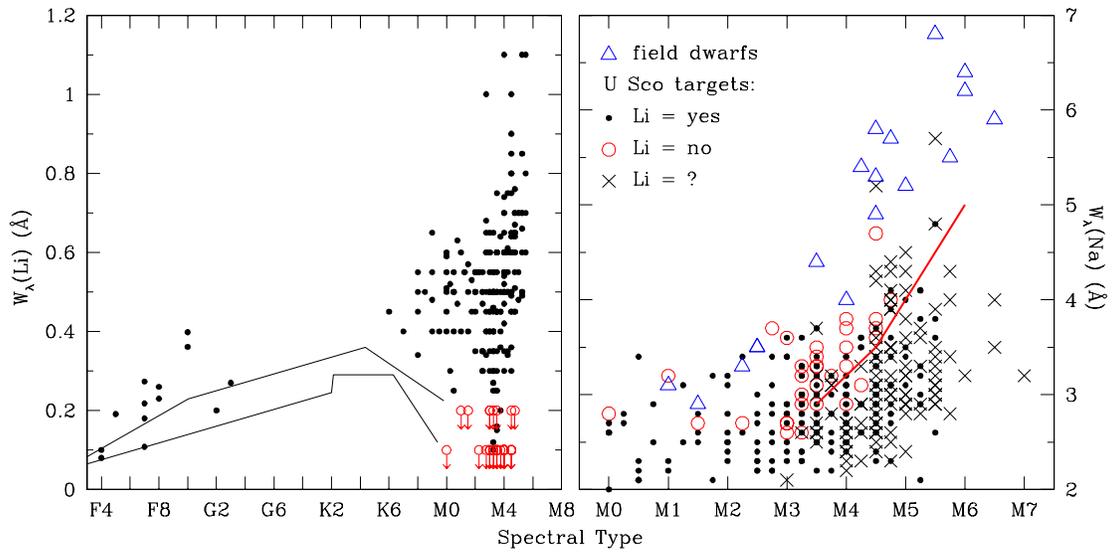}
\caption{
Equivalent widths of Li and Na versus spectral type for 
previously known and candidate members of Upper Sco for which we have
obtained optical spectra (Table~\ref{tab:spec}).
The left diagram contains detections of Li (filled circles) and
the useful upper limits ($<0.2$~\AA) that are available for non-detections
(open circles). In the right diagram, we show Na measurements for those
two samples and the stars that lack useful constraints on Li (crosses).
For comparison, we include the upper envelopes for Li data in IC~2602 (30~Myr)
and the Pleiades (125~Myr) \citep[upper and lower solid lines,][]{neu97}
in the left diagram and Na measurements for a sample of field dwarfs in the
right diagram. Most of these Li detections are consistent with the youth
expected for members of Upper Sco. The Na strengths for dwarfs and young,
Li-bearing stars diverge noticeably at $\gtrsim$M3.5. We take Na measurements
below the red curve as evidence of youth.
}
\label{fig:lina}
\end{figure}

We now describe how we used the Li absorption at 6707~\AA\ and the Na doublet
near 8190~\AA\ to assess the youth of our targets that were observed with
optical spectroscopy. Our measurements of the equivalent widths of these lines 
are included in Table~\ref{tab:spec}. Those data are reported only for dwarfs
and young stars, and the Na data are provided only for M types since they
are not useful as a gravity diagnostic at earlier types.
For some of the stars with optical spectra,
the signal-to-noise ratios were too low for useful measurements.
The typical strengths of Li and Na depend on both age and spectral type,
so we plot them as a function of spectral type in Figure~\ref{fig:lina}.
We also show upper envelopes for Li data in IC~2602 (30~Myr) and the
Pleiades (125~Myr) from \citet{neu97} and measurements of Na for a
sample of standard field dwarfs from our previous surveys and \citet{fil16}.
Li data are available from \citet{riz15} for a few stars that lack useful
Li constraints in our spectra, so we have adopted them.
In Figure~\ref{fig:lina}, most of the Li detections are stronger than expected 
for the two comparison clusters at 30 and 125~Myr, indicating ages that are 
consistent with that of Upper Sco \citep[11~Myr,][]{pec12,fei16}. The remaining 
detections fall along the upper envelopes for those clusters and could be 
consistent with Upper Sco membership given the uncertainties. Therefore, we 
take all of the Li detections in Figure~\ref{fig:lina} as evidence of
membership.

In Figure~\ref{fig:lina}, the standard field dwarfs exhibit stronger Na
than most of the young, Li-bearing stars in our spectroscopic sample,
which is a reflection of the dependence of this feature on surface gravity.
The two groups overlap at M0--M3 and diverge at later types.
The M stars in our sample with useful upper limits on their Li absorption
($<0.2$~\AA) have stronger Na on average than the Li-bearing stars, which
is consistent with the older ages implied by their Li constraints 
\citep[$\gtrsim20$~Myr,][]{cha97,dan97,sie00}.
To use Na for identifying targets that are likely to be young enough
to be members of Upper Sco, we
have defined a boundary in Figure~\ref{fig:lina} that is just below nearly
all of the stars at $\geq$M3.5 that lack Li strong absorption.
Measurements of Na below this boundary are taken as evidence of membership
unless Li absorption is absent.
Stars that appear above the boundary and that lack useful constraints on Li
can be considered possible members if they have other evidence of membership,
such as IR excess emission, or have noticeable reddening, which is not
expected for foreground field dwarfs.

In Table~\ref{tab:spec}, we list the spectral types and membership
classifications for the 768 objects in our spectroscopic sample.
Stars with membership flags of either ``Y" or ``Y?" are included in our
compilation of adopted members of Upper Sco in Table~\ref{tab:mem}.
Among the 530 spectroscopic targets that are adopted as
members, 377 have not been previously observed with spectroscopy.
Stars that have undetermined membership (marked as ``?") and that
satisfy our latest set of selection criteria from the CMDs and proper motions
are included in the sample of candidate members in Table~\ref{tab:cand}.
For the stars classified as members that have IR spectra, we have estimated
extinctions by comparing the observed spectral slopes to the
slopes of young standards from \citet{luh17}. For members that lack IR
spectra, we have derived extinctions from color excesses in $J-H$ relative
to the intrinsic photospheric values from \citet{ken95} and \citet{luh10}.
The companion HD~144218 has $K_s$ but not $J$, so $V-K_s$ was used instead.
The resulting extinction estimates in $K_s$ are included in Table~\ref{tab:mem}.
We have not attempted to estimate an extinction for Antares, which is
a supergiant, or the companion 2MASS~J16054012-2317155E, which lacks the
necessary data.

\subsection{Classifications of Previous Spectra}

In addition to classifying the objects in our spectroscopic sample, we have
measured new spectral types from IR spectra that have been obtained by 
previous studies for late-type ($\geq$M5) members of Upper Sco
\citep{luh05usco,lod08,lod18,all13,bow14,bow17,daw14,lac15,pen16,bes17}.
When deriving spectral types, some of these studies have assumed that the 
targets are not reddened by interstellar dust \citep[or have the same 
extinction as a more massive companion, as in the case of][]{bow17}.
However, according 
to our extinction estimates, 85, 10, and 4\% of the known members have
$A_K=0$--0.15, 0.15--0.3, and 0.3--0.5, respectively ($A_K\approx0.1$~$A_V$).
Thus, reddening was a free parameter in our classification of the previous
spectra, as done for our spectroscopic sample. In general, allowing for the
possibility of reddening results in a larger range of spectral types that
can match a spectrum. This is particularly true for L types since it is
primarily the spectral slope that varies with type rather than the depths of
the absorption bands in low-resolution data \citep[see Figure~17 in][]{luh17}.
Our new classifications are included in the compilation of spectral types 
for the known members in Table~\ref{tab:mem}.

\begin{figure}
\epsscale{1.2}
\plotone{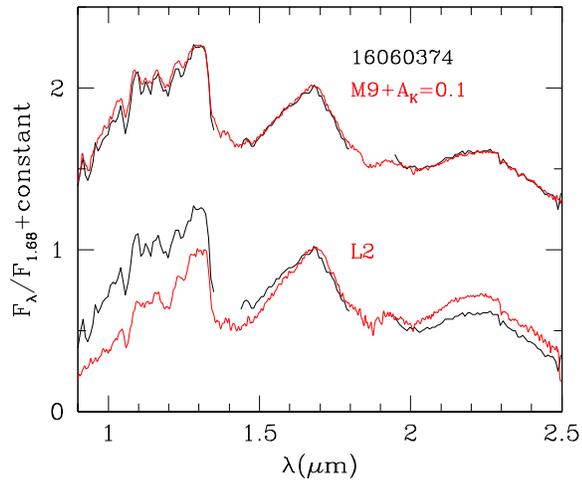}
\caption{
Example of a new reduction of an IR spectrum from \citet{lod08}
for a late-type member of Upper Sco. It has been binned to a resolution
of $R=150$. For comparison, we have included the young standard spectra
from \citet{luh17} for our adopted type of M9 and the type of L2 from
\citet{lod08}. The data used to create this figure are available.
}
\label{fig:ir2}
\end{figure}

\citet{lod08} did not reduce the two orders at 0.9--1.15~\micron\ in
their GNIRS data, so we have performed a new reduction to include them.
In Figure~\ref{fig:ir2}, we show an example of one of the newly reduced spectra,
which is compared to young standards from \citet{luh17} for our derived spectral
type and the type from \citet{lod08}. All of these newly reduced spectra are
provided in the electronic file that accompanies Figure~\ref{fig:ir2}.

\begin{figure}
\epsscale{1.2}
\plotone{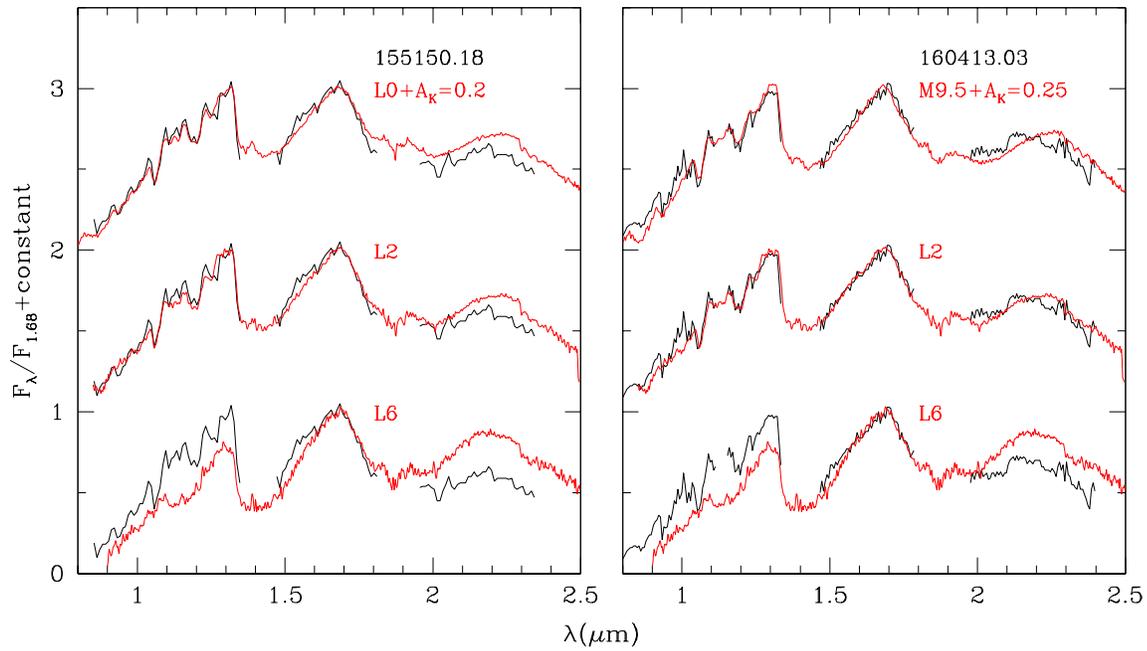}
\caption{
Examples of IR spectra of late-type members of Upper Sco from previous
studies \citep{pen16,lod18} for which we have measured new spectral types.
The data have been binned to a resolution of $R=150$.
The spectrum of each object is compared to two young standards from
\citet{luh17} that bracket our derived spectral type (top and middle) and 
the young standard from \citet{luh17} for the previously reported spectral
type (bottom).
}
\label{fig:ir3}
\end{figure}

In Figure~\ref{fig:ir3}, we show additional examples of previously reported
spectra that we have classified. The spectra are for
UGCS J155150.21$-$213457.2 and UGCS J160413.03$-$224103.2 \citep{pen16,lod18}.
We estimate spectral types of L0--L2 and M9.5--L2, respectively,
whereas \citet{pen16} and \citet{lod18} measured L6 for each of them.
In Figure~\ref{fig:ir3}, we compare each of the two spectra to the
standards from \citet{luh17} that bracket our classifications.
Among the L types, \citet{luh17} adopted young standard spectra for L0, L2,
L4, and L7. We have combined L4 and L7 with the appropriate weights to produce
L6, and the resulting spectrum has been included in Figure~\ref{fig:ir3}.
In our adopted classification system, UGCS J155150.21$-$213457.2 and
UGCS J160413.03$-$224103.2 are too blue to have types of L6.
A similar result was found for 1RXS J160929.1$-$210524~B in Figure~18
from \citet{luh17}. It was previously classified as L4, but it is much bluer
than our adopted L4 standard. Instead, \citet{luh17} found a good match to
M9.5 with $A_V=1.2$.

Most of our new classifications of previous IR spectra are earlier than
the originally reported types. One reason for this difference is that we have
accounted for the possibility of reddening from interstellar dust.
For instance, a reddened early L dwarf and an unreddened late L dwarf
can be indistinguishable in low-resolution spectra \citep{luh17}.
The choice of classification standards also contributes to the differences
between our spectral types and values from other studies.
In particular, the IR classifications from \citet{lod18} were largely based
on comparison to normal (i.e., not young) field L dwarfs and the Upper Sco
members 1RXS J160929.1$-$210524~B and J155150.21-213457.2, for which
they adopted the original types of L4 and L6.
However, the latter two objects have earlier types when
classified with our adopted standards, as we have discussed.
In addition, the most prominent near-IR spectral features of late-type objects
(H$_2$O bands and spectral slope) vary with surface gravity in such a way that
using older field dwarfs as the standards for classifying young objects
produces later types than if optically-classified young standards are used
\citep[][references therein]{luh12}.
Indeed, \citet{lod18} found that their IR classifications (which they adopted)
were systematically later than the optical types that they derived when
both IR and optical spectra were available. 
We prefer a classification scheme that produces the same spectral types,
on average, regardless of the wavelength range considered so that types
from different wavelengths can be utilized together in a meaningful way.

\begin{figure}
\epsscale{1}
\plotone{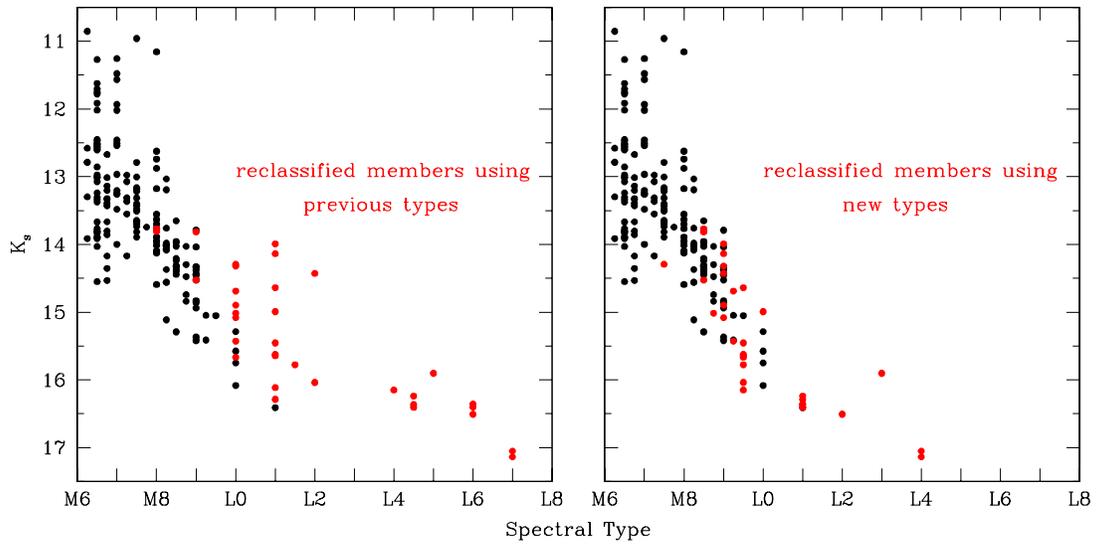}
\caption{
$K_s$ versus spectral type for known late-type members of Upper Sco.
We have measured new spectral types using IR spectra from previous studies
for the objects with red symbols. That sample is plotted with
the previous types (left) and our new types (right).
}
\label{fig:reclass}
\end{figure}

In Figure~\ref{fig:reclass}, we show two diagrams of $K_s$ versus spectral
type for the known members of Upper Sco that are later than M6.
For the objects that we have reclassified in this section,
we have used the previous types in the left diagram and our new types
in the right diagram. We have used the adopted types in Table~\ref{tab:mem}
for all other members in both diagrams. The sequence of members is tighter
and better defined when our new classifications are employed.

\section{Future Prospects for Upper Sco Census}

The census of Upper Sco compiled by \citet{luh12usco} contained
630 previously known members (excluding stars that we now reject).
Subsequent spectroscopic surveys have identified hundreds of additional
young stars that are likely to be members 
\citep[][this work]{daw14,riz15,pec16,pen16,lod18}.
Our latest sample of adopted members now contains 1631 objects.
Based on the large number of additional candidates that we have identified
($>1000$), substantial incompleteness likely remains in the current census
of the association. 
The second data release from {\it Gaia} should provide highly precise
measurements of proper motions and parallaxes down to substellar masses
in Upper Sco, which will greatly facilitate the identification of the
undiscovered members. The new {\it Gaia} data will also be valuable for
identifying nonmembers that are present in the current census.

A relatively complete census of Upper Sco, particularly in conjunction
with the astrometry from {\it Gaia}, will make it possible
to address the following questions more definitively than has been possible
previously:
What is the shape of the IMF in Upper Sco and how does it compare to
mass functions in other young nearby populations \citep{lod13c}?
How does the disk fraction vary with stellar mass and location
\citep{luh12usco}?
How does stellar age vary with stellar mass, disk class, and location
\citep{don17}?
Is a significant age spread present \citep{fan17}?
A full census of Upper Sco will also offer large, statistically
significant samples of targets for measurements of stellar multiplicity
and planet occurrence rates.

\acknowledgements
K.L. acknowledges support from NSF grant AST-1208239 and NASA grant
80NSSC18K0444. E.M. acknowledges support from the NASA NExSS program.
Part of this research was carried out at the Jet Propulsion Laboratory (JPL),
California Institute of Technology (Caltech), under a contract with NASA. 
The IRTF is operated by the University of Hawaii under contract NNH14CK55B
with NASA. The data at CTIO and SOAR were obtained through programs 
2011A-0415, 2013A-0224, 2014A-0180, 2015A-0192, and 2017A-0161 
at the National Optical Astronomy Observatory (NOAO).
CTIO and NOAO are operated by the Association of Universities for Research in
Astronomy under a cooperative agreement with the NSF. The Gemini data were 
obtained through programs GS-2009A-C-7 (NOAO 2009A-0083),
GN-2017A-FT-12, and GN-2017A-FT-17.
Gemini Observatory is operated by AURA under a cooperative agreement with
the NSF on behalf of the Gemini partnership: the NSF (United States), the NRC
(Canada), CONICYT (Chile), the ARC (Australia),
Minist\'{e}rio da Ci\^{e}ncia, Tecnologia e Inova\c{c}\~{a}o (Brazil) and
Ministerio de Ciencia, Tecnolog\'{i}a e Innovaci\'{o}n Productiva (Argentina).
{\it WISE} is a joint project of the University of California, Los Angeles,
and the JPL/Caltech, funded by NASA.
2MASS is a joint project of the University of Massachusetts and IPAC
at Caltech, funded by NASA and the NSF. This work used data from
the NASA/IPAC Infrared Science Archive, operated by JPL under contract
with NASA, and the SIMBAD database, operated at CDS, Strasbourg, France.
The DENIS project was partly funded by the SCIENCE and the HCM plans of
the European Commission under grants CT920791 and CT940627.
It was supported by INSU, MEN and CNRS in France, by the State of
Baden-W\"urttemberg in Germany, by DGICYT in Spain, by CNR in Italy, by
FFwFBWF in Austria, by FAPESP in Brazil, by OTKA grants F-4239 and F-013990
in Hungary, and by the ESO C\&EE grant A-04-046.
Jean Claude Renault from IAP was the Project manager. Observations were
carried out thanks to the contribution of numerous students and young
scientists from all involved institutes, under the supervision of P. Fouqu\'e.
The Pan-STARRS1 Surveys (PS1) and the PS1 public science archive have been made possible through contributions by the Institute for Astronomy, the University of Hawaii, the Pan-STARRS Project Office, the Max-Planck Society and its participating institutes, the Max Planck Institute for Astronomy, Heidelberg and the Max Planck Institute for Extraterrestrial Physics, Garching, The Johns Hopkins University, Durham University, the University of Edinburgh, the Queen's University Belfast, the Harvard-Smithsonian Center for Astrophysics, the Las Cumbres Observatory Global Telescope Network Incorporated, the National Central University of Taiwan, the Space Telescope Science Institute, the National Aeronautics and Space Administration under Grant NNX08AR22G issued through the Planetary Science Division of the NASA Science Mission Directorate, the NSF Grant AST-1238877, the University of Maryland, Eotvos Lorand University (ELTE), the Los Alamos National Laboratory, and the Gordon and Betty Moore Foundation.
This work has made use of data from the European Space Agency (ESA)
mission {\it Gaia} (\url{https://www.cosmos.esa.int/gaia}), processed by
the {\it Gaia} Data Processing and Analysis Consortium (DPAC,
\url{https://www.cosmos.esa.int/web/gaia/dpac/consortium}). Funding
for the DPAC has been provided by national institutions, in particular
the institutions participating in the {\it Gaia} Multilateral Agreement.
The Center for Exoplanets and Habitable Worlds is supported by the
Pennsylvania State University, the Eberly College of Science, and the
Pennsylvania Space Grant Consortium.

\clearpage

\begin{deluxetable}{ll}
\tabletypesize{\scriptsize}
\tablewidth{0pt}
\tablecaption{Members of Upper Sco\label{tab:mem}}
\tablehead{
\colhead{Column Label} &
\colhead{Description}}
\startdata
2MASS & 2MASS Point Source Catalog source name \\
WISEA & AllWISE Source Catalog source name\tablenotemark{a} \\
UGCS & UKIDSS Galactic Clusters Survey source name\tablenotemark{b}\\
Names & Other source names \\
RAdeg & Right Ascension (J2000) \\
DEdeg & Declination (J2000) \\
SpType & Spectral type \\
r\_SpType & Spectral type reference\tablenotemark{c} \\
Adopt & Adopted spectral type \\
Ak & Extinction in $K_s$ \\
f\_Ak & Method of extinction estimation\tablenotemark{d}
\enddata
\tablenotetext{a}{The following names are from the WISE All-Sky Source Catalog:
J160027.15-223850.5, J160414.16-212915.5, J161320.78-175752.3,
J161317.38-292220.0, J161837.22-240522.8, J162210.14-240905.4,
J162230.38-241119.2, J162620.15-223312.8.}
\tablenotetext{b}{Based on coordinates from Data Release 10 of the
UKIDSS Galactic Clusters Survey for stars with $K_s>10$ from 2MASS.}
\tablenotetext{c}{
(1) \citet{hou88};
(2) this work;
(3) \citet{kun99};
(4) \citet{pre98};
(5) \citet{pec16};
(6) \citet{tor06};
(7) \citet{esp18};
(8) \citet{riz15};
(9) \citet{daw14};
(10) measured in this work with the most recently published spectrum;
(11) \citet{lod06};
(12) \citet{lod08};
(13) \citet{bon14};
(14) \citet{rui87};
(15) \citet{bes17};
(16) \citet{hou82};
(17) \citet{vie03};
(18) \citet{pen16};
(19) \citet{pec12};
(20) \citet{cor84};
(21) \citet{shk09};
(22) \citet{all13b};
(23) \citet{ard00};
(24) \citet{wal94};
(25) \citet{mar10};
(26) \citet{kra09};
(27) \citet{mar04};
(28) \citet{sle08};
(29) \citet{pre02};
(30) \citet{mor01};
(31) \citet{rei08};
(32) \citet{kir10};
(33) \citet{all13};
(34) \citet{fah16};
(35) \citet{sle06};
(36) \citet{ria06};
(37) \citet{lod18};
(38) \citet{kra15};
(39) \citet{cod17};
(40) \citet{laf11};
(41) \citet{lac15};
(42) \citet{pre01};
(43) \citet{ans16};
(44) \citet{bej08};
(45) \citet{her09};
(46) \citet{kra07};
(47) \citet{lod11a};
(48) \citet{bil11};
(49) \citet{laf08};
(50) \citet{luh17};
(51) \citet{man16};
(52) \citet{dav16b};
(53) \citet{sta17};
(54) \citet{sta18};
(55) \citet{coh79};
(56) \citet{pra03};
(57) \citet{eis05};
(58) \citet{giz02};
(59) \citet{mar98a};
(60) \citet{pra07};
(61) \citet{cow69};
(62) \citet{lod15};
(63) \citet{dav16a};
(64) \citet{luh05usco};
(65) \citet{mar98b};
(66) T. Esplin, in preparation;
(67) \citet{gra06};
(68) \citet{lut77};
(69) \citet{cie10};
(70) \citet{bow11};
(71) \citet{bow14};
(72) \citet{rom12};
(73) \citet{jay06};
(74) \citet{clo07};
(75) \citet{luh07};
(76) \citet{mer10};
(77) \citet{mcc10};
(78) \citet{dav17};
(79) \citet{bra97};
(80) \citet{bou92};
(81) \citet{bow17};
(82) \citet{wil05};
(83) \citet{eri11};
(84) \citet{cie07}.}
\tablenotetext{d}{
Extinction estimated from a near-IR spectrum or the indicated color
assuming the intrinsic spectrum from \citet{luh17} or the intrinsic color
from \citet{ken95} and \citet{luh10} for the spectral type in question.}
\tablecomments{
The table is available in its entirety in machine-readable form.}
\end{deluxetable}

\begin{deluxetable}{ll}
\tabletypesize{\scriptsize}
\tablewidth{0pt}
\tablecaption{Candidate Members of Upper Sco\label{tab:cand}}
\tablehead{
\colhead{Column Label} &
\colhead{Description}}
\startdata
2MASS & 2MASS Point Source Catalog source name \\
WISEA & AllWISE Source Catalog source name \\
UGCS & UKIDSS Galactic Clusters Survey source name\tablenotemark{a}\\
RAdeg & Right Ascension (J2000) \\
DEdeg & Declination (J2000) \\
SpType & Spectral type \\
r\_SpType & Spectral type reference\tablenotemark{b} \\
Jmag & $J$ magnitude \\
e\_Jmag & Error in Jmag \\
Hmag & $H$ magnitude \\
e\_Hmag & Error in Hmag \\
Ksmag & $K_s$ magnitude \\
e\_Ksmag & Error in Ksmag \\
JHKref & $JHK_s$ reference\tablenotemark{c} \\
selection & Selection criteria satisfied by candidate\tablenotemark{d} 
\enddata
\tablenotetext{a}{Based on coordinates from Data Release 10 of the
UKIDSS Galactic Clusters Survey for stars with $K_s>10$ from 2MASS.}
\tablenotetext{b}{
(1) \citet{esp18};
(2) this work;
(3) \citet{hou88};
(4) \citet{sle08};
(5) \citet{hou82};
(6) \citet{riz15};
(7) \citet{cie07}.}
\tablenotetext{c}{
2 = 2MASS Point Source Catalog; u = UKIDSS Data Release 10; v = VISTA VHS Data
Release 5.}
\tablenotetext{d}{
G/W/i/Y/Z/ip/zp/yp = CMDs in Figure~\ref{fig:cmd} (all candidates also satisfy
the $H-K_s$ diagram in Figure~\ref{fig:cmd2});
pi = parallax from {\it Gaia} DR1;
gaia/gps/ucac/2m-gaia/2m-ps/ukidss = proper motions in Figure~\ref{fig:pm}.}
\tablecomments{
The table is available in its entirety in machine-readable form.}
\end{deluxetable}

\begin{deluxetable}{llll}
\tabletypesize{\scriptsize}
\tablewidth{0pt}
\tablecaption{Observing Log\label{tab:log}}
\tablehead{
\colhead{Telescope/Instrument\tablenotemark{a}} &
\colhead{Disperser/Aperture} &
\colhead{Wavelengths/Resolution} &
\colhead{Targets}}
\startdata
CTIO 4~m/Hydra & KPGLF/$2\arcsec$ fiber & 0.65--0.87~\micron/4~\AA & 289 \\
CTIO 4~m/COSMOS & red VPH/$0\farcs9$ or $1\farcs2$ slit & 0.55--0.95~\micron/3 or 4~\AA & 105 \\
CTIO 1.5~m/RC Spec & 600 \& 831 l~mm$^{-1}$/$2\arcsec$ slit & 0.37--0.69~\micron/3--4~\AA & 9 \\
SOAR/Goodman & 400 l~mm$^{-1}$/$0\farcs45$ slit & 0.54-0.94~\micron /3~\AA  & 74 \\
Gemini South/GMOS & R400/$0\farcs75$ slit & 0.57--0.99~\micron /6~\AA & 15 \\
Gemini North/GNIRS & 31.7~l~mm$^{-1}$/$1\arcsec$ slit & 0.9--2.5~\micron /R=600 & 8 \\
IRTF/SpeX & prism/$0\farcs8$ slit & 0.8--2.5~\micron/R=150 & 287
\enddata
\tablenotetext{a}{
The Gemini Multi-Object Spectrograph (GMOS), the Gemini Near-Infrared
Spectrograph (GNIRS), and SpeX are described by \citet{hoo04}, \citet{eli06},
and \citet{ray03}, respectively.
The Cerro Tololo Ohio State Multi-Object Spectrograph (COSMOS) is
based on an instrument described by \citet{mar11}.}
\end{deluxetable}

\begin{deluxetable}{lllllll}
\tabletypesize{\scriptsize}
\tablewidth{0pt}
\tablecaption{Spectroscopic Data for Previously Known and Candidate Members of Upper Sco\label{tab:spec}}
\tablehead{
\colhead{Source Name\tablenotemark{a}} &
\colhead{Spectral Type\tablenotemark{b}} &
\colhead{$W_{\lambda}$(Li)} &
\colhead{$W_{\lambda}$(Na)} &
\colhead{Instrument} &
\colhead{Date} &
\colhead{Member?}\\
\colhead{} &
\colhead{} &
\colhead{(\AA)} &
\colhead{(\AA)} &
\colhead{} &
\colhead{} &
\colhead{}}
\startdata
2MASS J15350863-2532397 & M5.25 & \nodata & \nodata & SpeX & 2012 Apr 23 & Y \\
2MASS J15355111-2021008 & M7.5 & \nodata & \nodata & SpeX & 2012 Apr 22 & Y \\
2MASS J15364206-2526186 & early G & 0.20 & \nodata & COSMOS & 2017 Jul 6 & Y? \\
2MASS J15374943-1920571 & G3 & 0.27 & \nodata & RC Spec & 2011 May 13/Jun 16 & N? \\
2MASS J15411302-2308161 & M5.5 & \nodata & \nodata & SpeX & 2012 Apr 24 & Y \\
2MASS J15413401-2507482 & M9 & \nodata & \nodata & SpeX & 2013 Jun 18 & Y 
\enddata
\tablenotetext{a}{Source names are from the 2MASS Point Source Catalog when
available. Otherwise, they are from the UKIDSS Galactic Clusters Survey.}
\tablenotetext{b}{Uncertainties are 0.25 and 0.5~subclass for optical and
IR spectral types, respectively, unless indicated otherwise.}
\tablecomments{
The table is available in its entirety in machine-readable form.}
\end{deluxetable}

\clearpage

\end{document}